\documentclass[]{mn2e}

\usepackage{graphics}
\usepackage{rotating}

\def\apj{ApJ}
\def\aap{A\&A}

\def\ueber#1#2{{\setbox0=\hbox{$#1$}%
  \setbox1=\hbox to\wd0{\hss$ #2$\hss}%
  \offinterlineskip
  \vbox{\box1\box0}}{}}

\def\lesssim{\,\lower 1mm \hbox{\ueber{\sim}{<}}\,}
\def\grsim{\,\lower 1mm \hbox{\ueber{\sim}{>}}\,}

\title{Simulating {\sl Gaia} performances on white dwarfs}

\author[Torres et al.]{Santiago Torres$^{1,2}$,
                       Enrique Garc\'{\i}a--Berro$^{1,2}$,
                       Jordi Isern$^{2,3}$ and
                       Francesca Figueras$^{2,4}$\\
           $^1$Departament de F\'\i sica Aplicada, Escola  Polit\`ecnica 
               Superior de Castelldefels, Universitat   Polit\`ecnica de 
               Catalunya, \\Avda.  del  Canal  Ol\'\i  mpic  s/n,  08860 
               Castelldefels, Spain\\
           $^2$Institute for Space Studies of Catalonia, c/Gran Capit\`a
               2--4, Edif.  Nexus 104, 08034 Barcelona, Spain\\
           $^3$Institut de  Ci\`encies de l'Espai, C.S.I.C., Campus UAB,  
               Facultat  de  Ci\`encies,  Torre  C-5,  08193 Bellaterra, 
               Spain\\
           $^4$Departament  d'Astronomia  i  Meteorologia, Universitat 
               de Barcelona,  Facultat  de  F\'\i sica,  Mart\'{\i}  i 
	       Franqu\`es 1, 08028 Barcelona, Spain}

\begin{document}

\maketitle

\begin{abstract}
One of  the most  promising space missions  of ESA is  the astrometric
satellite {\sl  Gaia}, which will provide very  precise astrometry and
multicolour photometry, for all 1.3 billion objects to $V\sim 20$, and
radial velocities with accuracies of  a few km s$^{-1}$ for most stars
brighter   than   $V\sim    17$.    Consequently,   full   homogeneous
six-dimensional  phase-space information  for a  huge number  of stars
will become  available.  Our  Monte Carlo simulator  has been  used to
estimate  the number of  white dwarfs  potentially observable  by {\sl
Gaia}.  From this we assess  which would be the white dwarf luminosity
functions  which {\sl  Gaia}  will  obtain and  discuss  in depth  the
scientific returns of {\sl Gaia}  in the specific field of white dwarf
populations.   Scientific attainable  goals include,  among  others, a
reliable  determination of  the age  of  the Galactic  disk, a  better
knowledge of the  halo of the Milky Way and  the reconstruction of the
star  formation  history  of  the  Galactic disk.   Our  results  also
demonstrate the potential  impact of a mission like  {\sl Gaia} in the
current understanding of the white dwarf cooling theory.
\end{abstract}


\section{Introduction}

{\sl  Gaia} is  an ambitious  space mission  (Perryman et  al.  2001),
adopted within  the scientific programme of the  European Space Agency
(ESA)    in   October    2000    ---   see,    for   instance,    {\tt
http://astro.estec.esa.nl/GAIA/}.   It aims  to measure  the positions
and  proper motions  of  an  extremely large  number  of objects  with
unprecedented accuracy.   As a result, a three-dimensional  map of the
positions (and velocities) of a  sizeable fraction of the stars of our
Galaxy  will  be  obtained,  as  well  as  solar  system  objects  and
extragalactic sources.  The precision of the angular measurements will
be of about 3~$\mu$as at 12$^{\rm mag}$, 10~$\mu$as at 15$^{\rm mag}$,
and  degrading to 200~$\mu$as  at 20$^{\rm  mag}$. The  satellite will
continuously scan  the sky,  allowing for astrometric  measurements on
the so-called Astrometric Field (AF), and for broad-band photometry on
a Broad  Band Photometer  (BBP).  Full sky  coverage will  be possible
because of the spin of the satellite around its own axis, which itself
precesses  at a  fixed angle  with respect  to the  Sun.   Spectra and
medium-band photometry  will also be obtained for  selected sources on
the SPECTRO  and MBP instruments, respectively, from  which the radial
velocity of the detected sources  will be obtained.  This will lead to
the most complete and accurate map of the stars of our Galaxy.

{\sl Gaia}  will be  the successor of  the astrometric  satellite {\sl
Hipparcos},  which was operative  from 1989  to 1993.   The scientific
program of  {\sl Hipparcos}  was more modest  than that of  {\sl Gaia}
since  it measured  the positions  and proper  motions of  only $10^5$
rather than  $\sim 10^9$ Galactic objects.   Moreover, {\sl Hipparcos}
operated on the basis of an input catalogue.  Instead, {\sl Gaia} will
determine its  own targets. In order to  do so, {\sl Gaia}  will use a
series  of sky  mappers  and a  sophisticated  on-board detection  and
selection algorithm  to produce a  list of targets to  be subsequently
followed  in the AF,  the MBP,  the BBP  and the  SPECTRO instruments.
Finally, the scientific products of {\sl Hipparcos} were released only
when the  mission was  complete, whereas some  of the  scientific data
that {\sl  Gaia} will  collect will be  partially released  during the
5~yr duration of the mission.

White  dwarfs  are  the  end-product  of the  evolution  of  low-  and
intermediate-mass stars. Thus, they preserve important clues about the
formation  and  evolution  of  our  Galaxy. This  information  can  be
retrieved by  studying their  observed mass, kinematic  and luminosity
distributions, provided that we  have good structural and evolutionary
models for  the progenitors  of white dwarfs  and for the  white dwarf
themselves.   In  particular,  the  fundamental tool  to  analyze  the
properties of the white dwarf population as a whole is the white dwarf
luminosity  function,  which  has  been consistently  used  to  obtain
estimates  of the  age  of the  Galactic  disk (Winget  et al.   1987;
Garc\'\i a--Berro  et al.   1988; Hernanz et  al. 1994; Richer  et al.
2000) and the past history of the Galactic star formation rate (Noh \&
Scalo 1990; D\'\i az--Pinto et al.   1994; Isern et al. 1995; Isern et
al. 2001). Although the situation  for the disk white dwarf population
seems to be relatively clear and well understood, this is not the case
for  the halo white  dwarf population.   The discovery  of microlenses
towards the Large Magellanic Cloud (Alcock et al.  1996, Alcock et al.
2000; Lasserre et  al.  2001) generated a large  controversy about the
possibility  that   white  dwarfs  could  be   responsible  for  these
microlensing   events   and,  thus,   could   provide  a   significant
contribution to the  mass budget of our Galactic  halo.  In both cases
{\sl Gaia} will potentially have  a large impact.  Consequently, it is
desirable to  foresee which  would be the  scientific returns  of {\sl
Gaia} in the field of white dwarfs.

In  this  paper we  assess  the  number  of white  dwarfs  potentially
observable by  {\sl Gaia}.   In doing this  our Monte  Carlo simulator
(Garc\'\i  a--Berro  et al.   1999;  Torres  et  al.  1998)  has  been
used. Our  results show which  could be the  impact of a  mission like
{\sl Gaia}  in the current  understanding of the Galactic  white dwarf
population.   This work  is organized  as follows.   In \S  2  a brief
description  of our  Monte Carlo  simulator  is given.   Section 3  is
devoted  to analyze  the  results of  our  simulations, including  the
completeness of  the samples  of disk and  halo white dwarfs,  and the
accuracy of  the astrometric determinations of both  samples.  In this
section a study  of the expected disk and  halo white dwarf luminosity
functions which  {\sl Gaia} will  presumably obtain is also  done, and
from  this the  scientific attainable  goals are  discussed  in depth.
Finally, \S 4  is devoted to summarize our  conclusions and to discuss
the results obtained here.

\section{The Monte Carlo simulator}

Since  our Monte  Carlo simulator  has been  thouroughly  described in
previous papers  (Garc\'\i a--Berro et al. 1999;  Garc\'\i a--Berro et
al. 2004) we will only summarize here the most important inputs. As in
any Monte Carlo simulation one  of the most important ingredients is a
random  number  generator.   We   have  used  a  pseudo-random  number
generator algorithm (James 1990)  which provides a uniform probability
density within the interval $(0,1)$ and ensures a repetition period of
$\grsim   10^{18}$,  which   is  virtually   infinite   for  practical
simulations. When  gaussian probability  functions are needed  we have
used   the   Box-Muller   algorithm   as   described   in   Press   et
al.  (1986).  Moreover,  each  one  of  the  Monte  Carlo  simulations
discussed below consists of an ensemble of 40 independent realizations
of the synthetic white dwarf  population, for which the average of any
observational quantity along with its corresponding standard deviation
were computed.  Here the standard deviation means the ensemble mean of
the sample dispersions for a typical sample.

In our simulations we have adopted a disk age ranging from 8 to 13 Gyr
(see  \S 3.3).   White dwarfs  have been  distributed according  to an
exponential  density   law  with  a   scale  length  $L=3.5$   kpc  in
Galactocentric radius.  A standard  initial mass function (Scalo 1998)
and a  constant volumetric star formation rate  were adopted, although
in  section \S 3.3  we also  explore other  star formation  rates. The
velocities have been drawn from normal laws:

\begin{eqnarray}
n(U)&\propto&{\rm e}^{-(U-U'_0)^2/\sigma^2_{\rm U}}\nonumber\\
n(V)&\propto&{\rm e}^{-(V-V'_0)^2/\sigma^2_{\rm V}}\\
n(W)&\propto&{\rm e}^{-(W-W'_0)^2/\sigma^2_{\rm W}}\nonumber
\end{eqnarray}

\noindent where $(U'_0,V'_0,W'_0)$  take into account the differential
rotation of the disk (Ogorodnikov  1965), and derive from the peculiar
velocity $(U_{\sun},V_{\sun},W_{\sun})$ of the sun with respect to the
local  standard  of  rest,  for   which  we  have  adopted  the  value
$(10,5,7)\;  {\rm km\; s^{-1}}$  (Dehnen \&  Binney 1997).   The three
velocity  dispersions  $(\sigma_{\rm  U},\sigma_{\rm  V},  \sigma_{\rm
W})$, and the lag velocity,  $V_0$, depend on the adopted scale height
(Mihalas \& Binney 1981):

\begin{eqnarray}
U_0&=&0\nonumber\\
V_0&=&-\sigma^2_{\rm U}/120\\
W_0&=&0\nonumber\\
&&\nonumber\\
\sigma^2_{\rm V}/\sigma^2_{\rm U}&=&0.32+1.67\ 10^{-5}\sigma^2_{\rm U}\nonumber\\
\sigma^2_{\rm W}/\sigma^2_{\rm U}&=&0.50\\
\sigma^2_{\rm W}&=&1.53\ 10^3 H_{\rm p}, \nonumber
\end{eqnarray}

\noindent where  the units are, respectively, kpc  and km~s$^{-1}$. We
have used $H_{\rm p}=500$~pc.   A standard model of Galatic absorption
has been used as well (Hakkila et al. 1997).

In the model the halo is assumed to be formed 14 Gyr ago in an intense
burst  of star  formation  of duration  1  Gyr (see,  however, \S  3.4
below).  The synthetic white dwarfs have been distributed according to
a typical isothermal, spherically symmetric halo:

\begin{equation}
\rho(r)=\rho_0\frac{a^2+R_{\sun}^2}{a^2+r^2}
\end{equation}

\noindent where $a\approx  5$ kpc is the core  radius, $\rho_0$ is the
local  halo  density  and  $R_{\sun}=$8.5 kpc  is  the  Galactocentric
distance of the  Sun. A standard initial mass  function was adopted as
well.  The  velocities of halo  stars were randomly drawn  from normal
distributions (Binney \& Tremaine 1987):

\begin{equation}
f(v_r,v_t)=\frac{1}{(2\pi)^{3/2}}\frac{1}{\sigma_r\sigma_t^2}
\exp\left[-\frac{1}{2}\left(\frac{v_r^2}{\sigma_t^2}+\frac{v_{t}^2}
{\sigma_t^2}\right)\right]
\end{equation}

\noindent  where $\sigma_r$  and  $\sigma_t$ ---  the  radial and  the
tangential velocity  dispersion, respectively  --- are related  by the
following expression:

\begin{equation}                                 
\sigma_t^2=\frac{V_{\rm c}^2}{2}+\left[1-\frac{r^2}{a^2+r^2}\right]
\sigma_r^2+\frac{r}{2}\frac{{\rm d}(\sigma_r^2)}{{\rm d}r}
\end{equation}

\noindent  which,  to  a  first  approximation,  leads  to  $\sigma_r=
\sigma_t=  {V_{\rm c}}/{\sqrt{2}}$  --- see,  for instance,  Binney \&
Tremaine (1987).  For the calculations reported here we have adopted a
circular  velocity $V_{\rm  c}= 220$~km/s.   From these  velocities we
obtain the heliocentric  velocities by adding the velocity  of the LSR
$v_{\rm LSR}=-220  $~km/s and the  peculiar velocity of the  sun.  The
velocity dispersions $\sigma_r$ and $\sigma_t$ are those determined by
Markovi\'c  \&  Sommer--Larsen   (1996).  In  particular,  the  radial
velocity dispersion is given by:

\begin{equation}
\sigma_r^2=\sigma_0^2+\sigma_+^2\left[\frac{1}{2}
-\frac{1}{\pi}\arctan\left(\frac{r-r_0}{l}\right)\right]
\end{equation}

\noindent whereas the tangential dispersion is given by:

\begin{equation}
\sigma_t^2=\frac{1}{2}V_{\rm c}^2-\left(\frac{\gamma}{2}-1\right)\sigma_r^2+
\frac{r}{2}\frac{{\rm d}\sigma_r^2}{{\rm d}r}
\end{equation}

\noindent where
\begin{equation}
r\frac{{\rm d}\sigma_r^2}{{\rm d}r}=
-\frac{1}{\pi}\frac{r}{l}\frac{\sigma_+^2}{1+[(r-r_0)/l]^2}
\end{equation}

\noindent The values of the constants are, respectively $\sigma_0=80\,
{\rm  km\,s^{-1}}$, $\sigma_+=145\, {\rm  km\,s^{-1}}$, $r_0=10.5$~kpc
and $l=5.5$~kpc,

The procedure to  obtain the synthetic stars is  the following. First,
we randomly  choose the three-dimensional coordinates of  each star of
the sample according to the  adopted density laws.  Afterwards we draw
another pseudo-random number in order to obtain the main sequence mass
of each star,  according to the initial mass  function.  Once the mass
of the progenitor  of the white dwarf is known  we randomly choose the
time at which each star was born, according to the star formation rate
of the  population under  study.  Given the  age of  the corresponding
population and the main sequence lifetime as a function of the mass in
the main  sequence (Iben  \& Laughlin 1989)  we know which  stars have
lived long  enough to become  white dwarfs, and  given a set  of fully
evolutionary cooling sequences for several white dwarf masses (Salaris
et  al.  2000)  --- which  reproduce  the so-called  ``blue hook''  of
hydrogen-rich  (DA) white  dwarfs ---  and the  initial to  final mass
relationship (Iben  \& Laughlin 1989), their  present day luminosities
and  magnitudes.   The  magnitude   is  then  corrected  for  Galactic
extinction and  reddening and converted to  the instrumental magnitude
of {\sl Gaia}, $G$, which is related to the standard colours $(V,V-I)$
by the expression (Perryman 2002):

\begin{eqnarray}
G-V =&-&0.00544 -0.36919 (V-I) -0.09727 (V-I)^2\nonumber \\
     &+&0.00372 (V-I)^3
\end{eqnarray}

\noindent  The typical  errors  in parallax  depend  on the  magnitude
(Perryman 2002) and are computed using the following fitting function:

\begin{eqnarray}
\sigma_{\pi}&\simeq&\sqrt{7+105z+1.3z^2+6 \,10^{-10}z^6}\nonumber \\
                  & &\mbox{}\times\big[0.96+0.04\,(V-I)\big]
\end{eqnarray}

\noindent   where  $\sigma_\pi$   is  given   in  $\mu$as   and  $\log
z=0.4(G-15)$.   The  errors  in  parallax,  $\sigma_\pi$,  and  proper
motion,   $\sigma_{\mu}$,  in   $\mu$as~yr$^{-1}$,   are  related   by
$\sigma_{\mu} =0.75\sigma_{\pi}$ (Perryman 2002).
 
\begin{figure}
\vspace{7cm}
\includegraphics{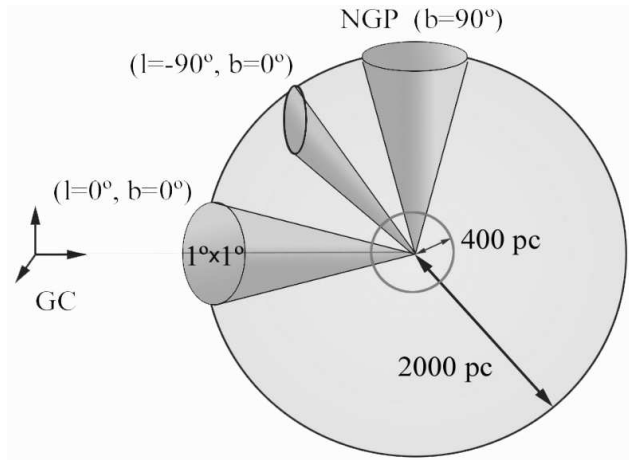}
\caption{The adopted geometry for the Monte Carlo simulations.}
\label{geometry}
\end{figure}

In this way  we end up with all the  relevant information necessary to
assess the performance  of {\sl Gaia}.  It is  important to mention at
this  point that  since the  number  of synthetic  stars necessary  to
simulate  the  whole  Galaxy  is  prohibitively  large  we  have  only
determined  the  number  of  white  dwarfs that  could  eventually  be
detected within 400~pc of the sun.  The number density of white dwarfs
in this sphere  of radius 400~pc was normalized  to the local observed
density  of either disk  (Oswalt et  al.  1996)  or halo  white dwarfs
(Torres  et  al.  1998),  depending  on  the  population under  study.
Additionally, we have  computed the total number of  white dwarfs in a
small window of $1^{\circ}\times 1^{\circ}$ (termed a ``pencil beam'')
for  each  of the  three  directions  shown in  figure~\ref{geometry}.
Since  the  direction  $l=0^\circ$,  $b=0^\circ$  corresponds  to  the
direction  in  the  Galactic  plane  for  which  Galactic  extinction,
statistically  speaking, is  expected to  be  a maximum  we have  also
computed  the   total  number  of   white  dwarfs  in   the  direction
$l=180^\circ$, $b=0^\circ$, which  corresponds to a minimum extinction
on the  Galactic plane.   Each of  these pencil beams  has a  depth of
2000~pc and is again normalized to the observed local density of white
dwarfs.  Obviously,  only the brightest  white dwarfs will be  seen at
very large  distances.  After doing  this we average our  results over
the corresponding octant  and we repeat the procedure  for the rest of
octants.  Finally we  average our results over the  whole sky in order
to obtain an estimate of the total number of white dwarfs within 2~kpc
of the sun accessible to {\sl Gaia}.

\section{Results}

\subsection{Expected number of disk and halo white dwarfs} 

\begin{figure}
\vspace{13cm}
\includegraphics{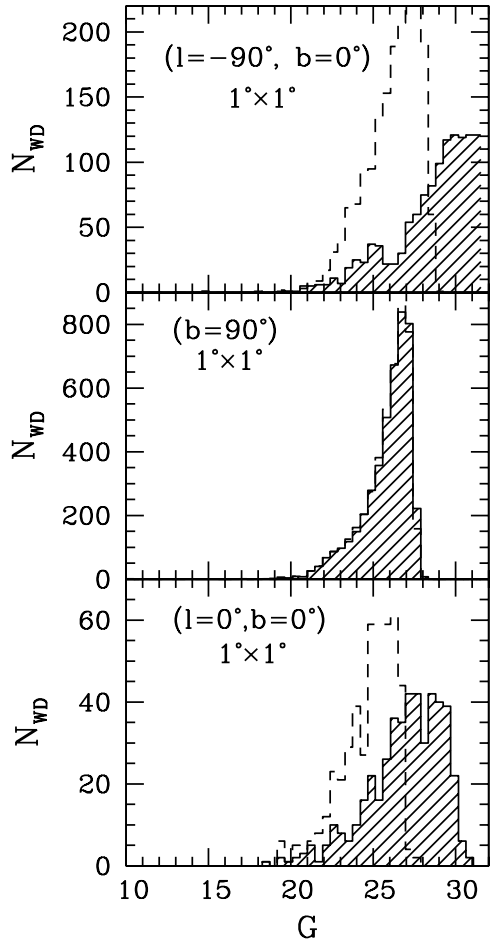}
\caption{Total number of  synthetic white dwarfs as a  function of the
         {\sl Gaia}  apparent magnitude along the  three small pencils
         beams   discussed  in  the   text.   The   shaded  histograms
         correspond   to  the   population   obtained  when   Galactic
         extinction is  taken into  account, whereas the  dashed lines
         correspond   to  the   population   obtained  when   Galactic
         extinction is neglected in the calculations.}
\label{pencils}
\end{figure}

The total number of disk  white dwarfs along the three above mentioned
pencil beams of the first octant  and the total number of white dwarfs
accessible to {\sl  Gaia} are shown in table  1 for different limiting
magnitudes, between  $G_{\rm lim}=20$ and 21, since  the real limiting
magnitude of {\sl Gaia} will  depend on the specific technical design.
Also, in figure~\ref{pencils} we show the distribution of white dwarfs
in a typical realization of the Monte Carlo simulations, as a function
of their apparent $G$ magnitude for each one of the three pencil beams
considered  here,   and  $G_{\rm  lim}=20$.    The  shaded  histograms
correspond to  the case  in which Galactic  extinction has  been taken
into account, whereas the dashed  histograms correspond to the case in
which  Galactic  extinction  was  disregarded.   As  can  be  seen  in
figure~\ref{pencils}  the  effects of  Galactic  extinction are  quite
evident.  Note  as well  that the number  of white  dwarfs potentially
observable  by   {\sl  Gaia}  decreases   considerably  when  Galactic
extinction is taken into account.

\begin{table}
\begin{center}  
\caption{Number  of disk white  dwarfs for  three patches  of $1^\circ
         \times 1^\circ$ in different regions of the sky (see text for
         details).}
\begin{tabular}{lccc}  
\hline  
& $G<19$ & $G<20$ &  $G<21$ \\  
\hline  
$(l=-90^{\circ},\, b=0^{\circ})$  &  6 &  9 & 13 \\  
$(l=0^{\circ},\,   b=0^{\circ})$  &  5 & 11 & 13 \\
$(l=180^{\circ},\, b=0^{\circ})$  &  6 & 11 & 15 \\
$(b=90^{\circ})$                  &  4 &  8 & 12 \\  
\hline
All sky  & $2.1\cdot10^5$ &  $3.9\cdot 10^5$  & $5.2\cdot 10^5$ \\ 
\hline 
\end{tabular} 
\end{center} 
\end{table}

\begin{figure*}
\vspace{10.7cm}
\includegraphics{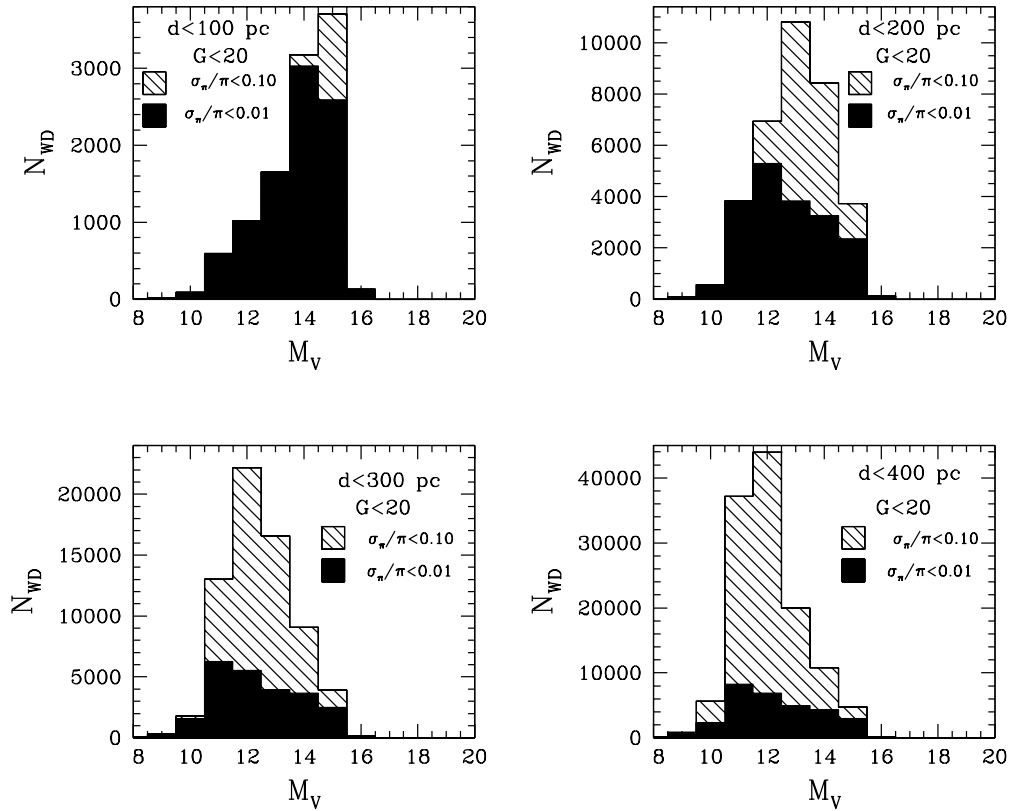}
\caption{Distribution of  the number of disk white  dwarfs detected by
         {\sl  Gaia}  as  a   function  of  their  absolute  magnitude
         according  to   their  errors  in   parallax,  for  different
         distances.}
\label{ndisk}
\end{figure*}

In figure~\ref{ndisk} we  show the distribution in the  number of disk
white dwarfs  detected by {\sl Gaia}  as a function  of their absolute
magnitude, according  to their errors  in parallax, for 100,  200, 300
and  400 pc and  $G_{\rm lim}=20$.   Fig.~\ref{ndisk} shows  that {\sl
Gaia} will detect faint white dwarfs up to considerable distances.  It
is noticeable  that {\sl Gaia}  will detect white dwarfs  with $M_{\rm
v}\simeq 16^{\rm  mag}$ --- that of  the observed cut-off  of the disk
white dwarf luminosity function --- up to distances of 100~pc.

The total  estimated number of white  dwarfs within 100,  200, 300 and
400  pc can  be found  in the  first row  of table  2. Of  these white
dwarfs,  those which  pass the  cut in  $G$ apparent  magnitude  for a
limiting magnitude of 21 are given  in the second row of table 2.  The
third  row lists  which  of  these will  also  have measurable  proper
motions. We  have considered that  a white dwarf will  have measurable
proper  motion when the  error in  proper motion  is smaller  than the
proper  motion itself.   The completeness  ($\nu_{21}$) of  the sample
necessary to build the white  dwarf luminosity function is assessed in
the last row of  this table.  As can be seen, {\sl  Gaia} will be able
to detect the whole white  dwarf population within 100~pc, and roughly
half  of it within  300~pc, decreasing  to one  third at  distances of
400~pc, which represents  a huge step forward in  our knowledge of the
white dwarf population.  The second  and the third sections of table 2
assess  the accuracy  of the  astrometric measurements.   Most  of the
detected  white dwarfs  will  have good  determinations  for both  the
parallax and the  proper motion ($\sigma_\mu/\mu$ and $\sigma_\pi/\pi<
0.1$)  up to  distances  of more  than  400 pc  and superb  accuracies
($\sigma_\mu/\mu$  and $\sigma_\pi/\pi<  0.01$) will  be  obtained for
half of the sample up to  distances of about 200 pc. The last sections
of table  2 list the same  quantities for a limiting  magnitude of 20.
Obviously the overall performances  and the completeness of the sample
will be in this case smaller.

\begin{table}
\centering
\caption{Results of  the Monte Carlo  simulations of disk  white dwarf
         population accessible to {\sl Gaia}.}
\vspace{0.3cm}
\begin{tabular}{lcccc}
\hline
\hline
\  & 100 pc & 200 pc & 300 pc & 400 pc \\
\hline
$N_{\rm WD} (9.0<G<29.0)$ &  11595 & 80775 &  273684 &  743893\\
\hline
$N_{\rm WD} (G<21)$       &  11593 & 57804 &  121474 &  226922\\
$N_{\rm WD} (\mu>\mu_{\rm cut})$ & 11593 & 57804 &  121472 & 226919\\
\hline
$\sigma_{\mu}/\mu<0.10$ & 1.000 & 0.999 & 0.999 &  0.998 \\
$\sigma_{\mu}/\mu<0.01$ & 0.998 & 0.966 & 0.929 &  0.893 \\
\hline
$\sigma_{\pi}/\pi<0.10$ & 1.000 & 1.000 & 1.000 &  1.000 \\
$\sigma_{\pi}/\pi<0.01$ & 0.877 & 0.559 & 0.357 &  0.249 \\
\hline
$\nu_{21}$ & 0.999 & 0.716 & 0.444 &  0.305\\
\hline
$N_{\rm WD} (G<20)$       &  10391 & 34560 &   67067 &  123419\\
$N_{\rm WD} (\mu>\mu_{\rm cut})$ & 10391 & 34560 &  67067 & 123418\\
\hline
$\sigma_{\mu}/\mu<0.10$ & 1.000 & 0.999 & 0.999 &  0.999 \\
$\sigma_{\mu}/\mu<0.01$ & 0.998 & 0.994 & 0.986 &  0.976 \\
\hline
$\sigma_{\pi}/\pi<0.10$ & 1.000 & 1.000 & 1.000 &  1.000 \\
$\sigma_{\pi}/\pi<0.01$ & 0.877 & 0.559 & 0.357 &  0.249 \\
\hline
$\nu_{20}$ & 0.896 & 0.428 & 0.245 &  0.166\\
\hline
\hline
\end{tabular}
\end{table}

\begin{table}
\centering
\caption{Same as table 2 for the halo white dwarf population.}
\vspace{0.3cm}
\begin{tabular}{lcccc}
\hline
\hline
\  & 100 pc & 200 pc & 300 pc & 400 pc \\
\hline
$N_{\rm WD} (9.0<G<29.0)$ &  359 & 2737 &  9174 &  21505 \\
\hline
$N_{\rm WD} (G<21)$       &  192 & 434  &  726 &  1099\\
$N_{\rm WD} (\mu>\mu_{\rm cut})$ & 192 & 434 &  726 & 1099\\
\hline
$\sigma_{\mu}/\mu<0.10$ & 1.000 & 1.000 & 1.000 &  1.000 \\
$\sigma_{\mu}/\mu<0.01$ & 1.000 & 1.000 & 0.998 &  0.995 \\
\hline
$\sigma_{\pi}/\pi<0.10$ & 1.000 & 1.000 & 1.000 &  1.000 \\
$\sigma_{\pi}/\pi<0.01$ & 0.905 & 0.636 & 0.395 &  0.260 \\
\hline
$\nu_{21}$  &  0.535 & 0.158 & 0.079 &  0.051\\
\hline
$N_{\rm WD} (G<20)$       &   84 &  195 &   344 &   542\\
$N_{\rm WD} (\mu>\mu_{\rm cut})$ &  84 & 195 &  344 &  542\\
\hline
$\sigma_{\mu}/\mu<0.10$ & 1.000 & 1.000 & 1.000 &  1.000 \\
$\sigma_{\mu}/\mu<0.01$ & 1.000 & 1.000 & 0.998 &  1.000 \\
\hline
$\sigma_{\pi}/\pi<0.10$ & 1.000 & 1.000 & 1.000 &  1.000 \\
$\sigma_{\pi}/\pi<0.01$ & 0.905 & 0.636 & 0.395 &  0.260 \\
\hline
$\nu_{20}$  &  0.234 & 0.071 & 0.038 &  0.025\\
\hline\hline
\end{tabular}
\end{table}

\begin{figure*}
\vspace{10.5cm}
\includegraphics{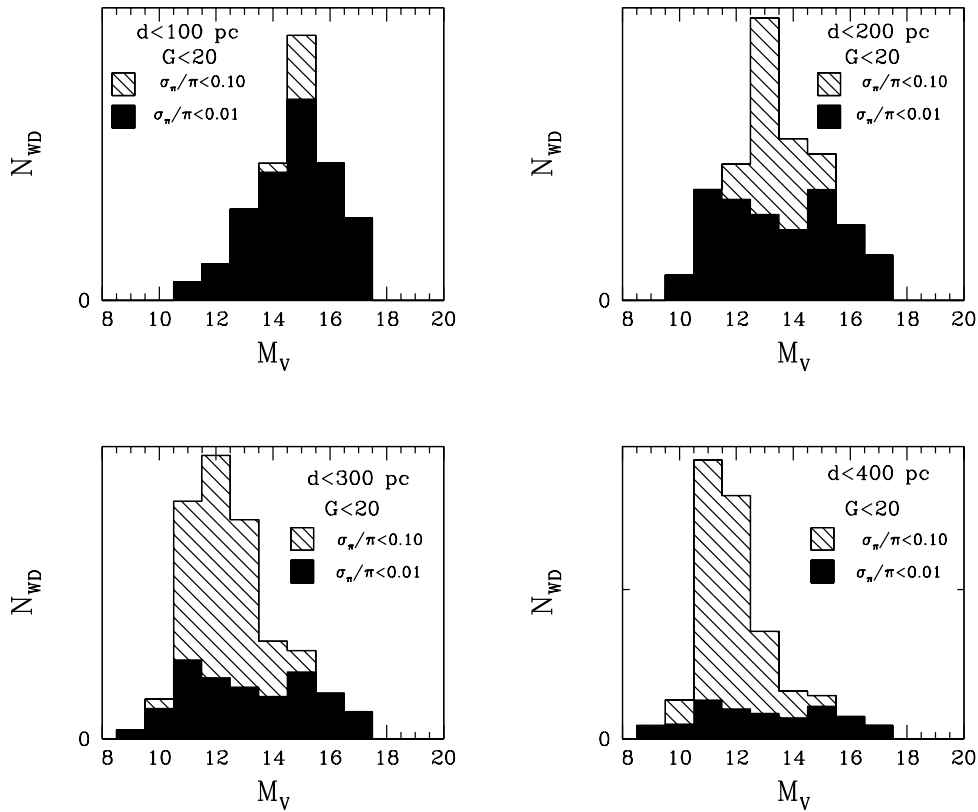}
\caption{Same as figure \ref{ndisk} for the halo white dwarf population.}
\label{nhalo}
\end{figure*}

The same  exercise can  be done for  the halo white  dwarf population.
However the  results are not as  encouraging as those  obtained so far
for the disk  white dwarf population.  The results  for the halo white
dwarf  population are  displayed  in table  3 and  figure~\ref{nhalo}.
Perhaps the most  important result of this set  of simulations is that
the  number of  halo white  dwarfs which  {\sl Gaia}  will be  able to
observe  is  likely  to be  of  the  order  of  a few  hundreds,  thus
increasing enormously the total number of halo white dwarf candidates,
which presently is of order  10 or less.  However, the completeness of
the  sample even  for  100 pc  will be  small  --- $\sim  50$\% for  a
limiting magnitude of 21 and only $\sim 20$\% for a limiting magnitude
of  20  ---  and   decreasing  rapidly  for  larger  distances.   Most
importantly,  {\sl Gaia}  will  only  be able  to  observe the  bright
portion of the halo white  dwarf luminosity function.  Hence, a direct
determination of  the age of  the halo using  the cut-off of  the halo
white  dwarf luminosity function  will not  be possible.   Despite all
this,  the accuracy  of the  measurements of  those halo  white dwarfs
detected by {\sl Gaia} will be impressive since we will have extremely
precise  parallaxes for  a good  fraction  of halo  white dwarfs  with
distances of up to 400 pc, independently of the limiting magnitude.

\subsection{Classifying halo and disk white dwarfs}

The natural question which arises now is how to distinguish halo white
dwarfs  from  disk  white  dwarfs.  Obviously,  due  to  gravitational
settling in the atmospheres of  white dwarfs the metallicity cannot be
used. Although  {\sl Gaia}  will be able  to obtain  radial velocities
using  the SPECTRO  instrument it  is unlikely  that {\sl  Gaia} could
determine the full three-dimensional velocities of white dwarfs, since
the  SPECTRO instrument  will be  optimized for  $G<17$  main sequence
stars.   The reduced proper  motion diagram  can be  of great  help in
distinguishing  disk and  halo members.   An example  of  the expected
reduced proper motion diagram that  {\sl Gaia} will obtain is shown in
figure~\ref{Hdh}.   As can be  seen there,  the reduced  proper motion
$H=M_{\rm V}-5\log\pi+5\log\mu$ is a  good indicator of the membership
to  a given  population. Simulated  halo white  dwarfs occupy  a clear
locus in  this diagram.  However, for $V-I\la  0.2$ the identification
becomes  less  clear.  Also  shown  in  figure~\ref{Hdh}  is the  line
$H_0=7(V-I)+17.2$.   Clearly,   white  dwarfs  with   $H>H_0$  can  be
preliminarily ascribed  to the halo white dwarf  population.  In order
to make quantitative statements we introduce the confusion matrix:

\begin{equation}
C_H=
\left(
\begin{array}{cc}
0.68 &0.05\\
0.32 &0.95\\
\end{array}
\right)
\end{equation}

\noindent where the matrix element $C_H^{11}$ indicates the percentage
of  disk  white  dwarfs  classified  as ``disk'',  $C_H^{21}$  is  the
percentage of disk stars missclassified as ``halo'', and so on. Hence,
the  reduced  proper  motion,  $H$,  turns  out  to  be  a  reasonable
membership discriminator.

In order to provide a more consistent and easy method to ascertain the
population  membership of  white dwarfs  we proceed  in the  spirit of
Salim \& Gould  (2003). That is, we introduce  a discriminator $\eta$,
by which  we classify  stars as  a function of  their position  in the
reduced  proper  motion  diagram,  their colour,  and  their  Galactic
latitude:

\begin{equation}
\eta=H+C_1(V-I)+C_2 |\sin b| + C_3
\end{equation}

\noindent where the  constants $C_1$, $C_2$ and $C_3$  are computed in
such a way that the distance, $s$, between the halo and the disk white
dwarf populations is maximum:

\begin{equation}
s=\sum_i^{N_{\rm D}} \sum_j^{N_{\rm H}} (\eta_i-\eta_j)^2
\end{equation}

\noindent  where $N_{\rm D}$  and $N_{\rm  H}$ are,  respectively, the
number of simulated disk and halo white dwarfs.

The  value of the  constants obtained  in such  a way  are $C_1=0.48$,
$C_2=0.40$  and  $C_3=-4.83$.   This  membership  discriminator  works
slightly  better  than the  reduced  proper  motion and  distinguishes
between disk and  halo white dwarfs, but, again,  only for those white
dwarfs with colour indices $V-I\ga 0.2$.  Moreover, those white dwarfs
with $\eta > \eta_0=3 (V-I)+ 13.1$ can be considered as good bona-fide
halo white dwarfs.  It is  interesting to realize that less disk white
dwarfs are misclassified as halo  members and vice versa.  In fact the
confusion matrix is in this case:

\begin{figure}
\vspace{6.5cm}
\includegraphics{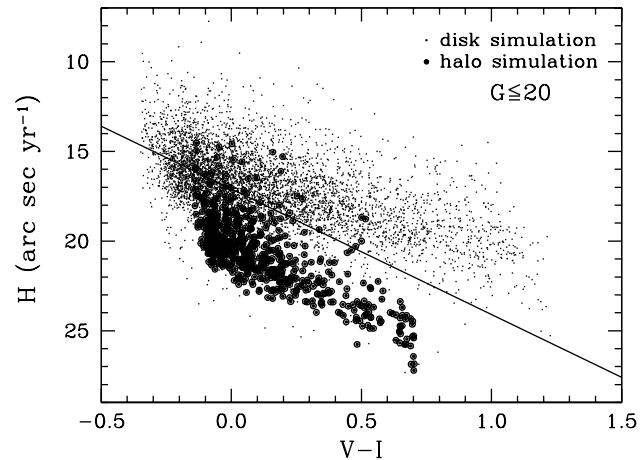}
\caption{Reduced proper  motion diagram for  the disk ---  small solid
         dots   ---   and   halo    ---   large   open   circles   ---
         simulations. Only those white dwarfs with $m_{\rm G}<20$ have
         been  considered. For  the sake  of clarity  only 5\%  of the
         simulated disk white dwarfs have been plotted.}
\label{Hdh}
\end{figure}

\begin{figure}
\vspace{6.5cm}
\includegraphics{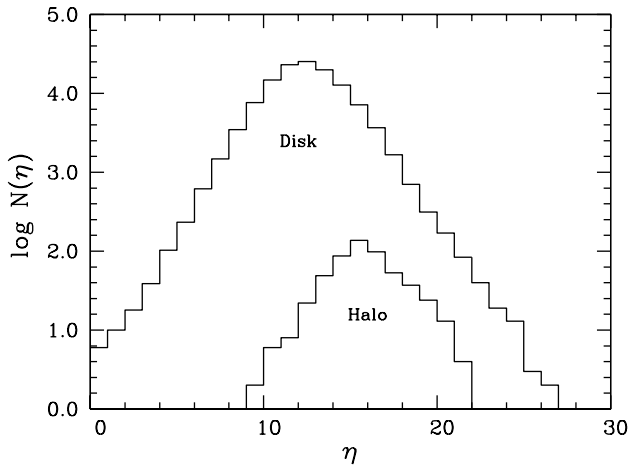}
\caption{Distribution  of  the   membership  discriminator  $\eta$.  A
         logarithmic scale has been used for the sake of clarity.}
\label{Heta}
\end{figure}

\begin{equation}
C_\eta=
\left(
\begin{array}{cc}
0.73 &0.06\\
0.27 &0.94\\
\end{array}
\right)
\end{equation}

Figure~\ref{Heta} demonstrates the difficulty of recovering halo white
dwarfs.  In  this figure  we show the  distribution of  our membership
discriminator for both the halo and disk white dwarf populations. Both
distributions  are  shown in  a  logarithmic  scale  for the  sake  of
clarity.  By examining this figure  it turns out that the distribution
of halo white dwarfs cannot  be clearly discriminated from that of the
disk.   Moreover,  disk white  dwarfs  outnumber  halo members.   Both
distributions are  approximately gaussian, and the  average values and
standard deviations  of the membership discriminator for  the disk and
halo  white dwarf  populations turn  out to  be $\langle  \eta_{\rm D}
\rangle \simeq  12.4\pm 2.3$ and $\langle \eta_{\rm  H} \rangle \simeq
16.1\pm 2.2$, respectively.   Using this membership discriminator 82\%
and  83\% of  the  whole disk  and  halo white  dwarf populations  are
correctly  classified  at the  $1\sigma$  level.   It  must be  noted,
however,  that an  artificial  intelligence algorithm  (Torres et  al.
1998;  Garc\'\i a--Berro  et al.   2003a) can  be succesfully  used to
classify  white  dwarfs, and  that  the  results  obtained using  that
algorithm are considerably better, since only 2\% of disk white dwarfs
are  erronously  classified  as   halo  white  dwarfs  using  advanced
classification techniques.

\subsection{The disk white dwarf luminosity function}

\begin{figure}
\vspace{6.5cm}
\includegraphics{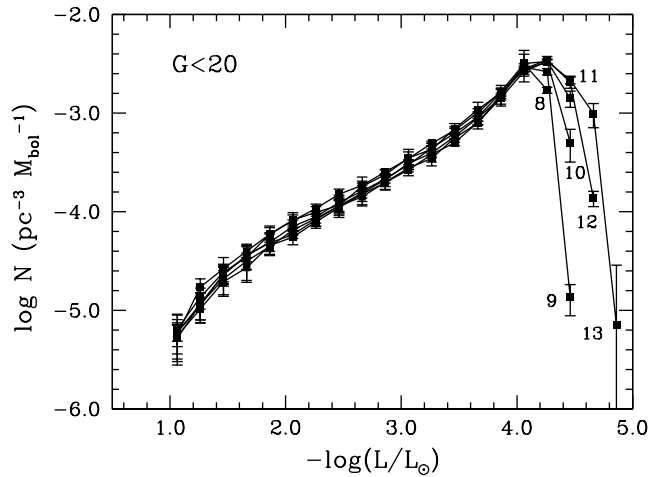}
\caption{Luminosity function of disk  white dwarfs for several ages of
         the  disk, ranging  from 8  to  13~Gyr, with  an interval  of
         1~Gyr. The  error bars are  the standard deviation of  the 40
         independent realizations. See text for details.}
\label{fldiskage}
\end{figure}

Now that we have assessed the total number counts of disk white dwarfs
we pay  attention to  some specific matters  regarding the  disk white
dwarf luminosity function. In particular  we ask to what precision the
age of the Galactic disk can be estimated.  In Fig.~\ref{fldiskage} we
show  the average  of 40  independent realizations  of the  disk white
dwarf  luminosity  function for  several  disk  ages.   Each curve  is
labelled with its  corresponding age. The error bars  are the standard
deviation  of  the  40  independent  realizations.   The  white  dwarf
luminosity  functions have  been  computed using  the $1/V_{\rm  max}$
method (Schmidt  1968).  Hence,  a set of  restrictions is  needed for
selecting  a subset  of white  dwarfs which,  in principle,  should be
representative of the whole white dwarf population. We have chosen the
following criteria for selecting the final sample: $G\le 20^{\rm mag}$
and no  restriction in proper motion.   The reason for  this choice is
quite simple.   From the  discussion in  \S 3.1 it  is clear  that the
sample of disk white dwarfs  that {\sl Gaia} will eventually detect is
almost complete in magnitude up to $G \simeq 20$, and all white dwarfs
within this sample will have measurable proper motions.  Consequently,
the proper motion cut does not play  any role at all. This is in sharp
contrast with the adopted magnitude and proper motion cuts actually in
use  to derive  the  observed disk  white  dwarf luminosity  function:
$m_{\rm  V}\le 18.5^{\rm  mag}$ and  $\mu\ge 0.16^{\prime\prime}\;{\rm
yr}^{-1}$ (Oswalt  et al.  1996).  Additionally, and  since the number
of  white  dwarfs that  are  used in  building  the  disk white  dwarf
luminosity function is much larger  than the size of the actual sample
of  white dwarfs  with known  parallaxes  and proper  motions we  have
binned  the luminosity  function  in smaller  luminosity  bins. To  be
precise, the binning is five bins per decade.

\begin{table}
\centering
\caption{Expected statistical  errors in the determination  of the age
         of the disk, as obtained from fitting the cut-off in the disk
         white dwarf luminosity  function, in terms of the  age of the
         disk. See text for details.}
\vspace{0.3cm}
\begin{tabular}{cc}
\hline
$T_{\rm disk}$~(Gyr) & $\Delta T_{\rm disk}$~(Gyr) \\
\hline
 8 & 0.15\\
 9 & 0.30\\
10 & 0.30\\
11 & 0.30\\
12 & 0.15\\
13 & 1.13\\
\hline
\end{tabular}
\end{table}

Fig.~\ref{fldiskage}  clearly  shows that  the  error  bars are  small
enough  to secure a  reliable determination  of the  age of  the disk,
using  the  observed  cut-off  in  the  disk  white  dwarf  luminosity
function.   The easiest  and more  straightforward way  to  assess the
statistical errors associated  with the measurement of the  age of the
disk is trying to reproduce  the standard procedure.  That is, we have
fitted the  position of the  ``observational'' cut-off of each  of the
Monte Carlo realizations with a  standard method (Hernanz et al. 1994)
to compute the white dwarf luminosity function using {\sl exactly} the
same inputs adopted to  simulate the Monte Carlo realizations, except,
of course, the age of the  disk, which is the only free parameter.  To
be more precise,  we compute the disk white  dwarf luminostiy function
according to:

\begin{eqnarray}
n(L)&\propto&\int^{M_{\rm s}}_{M_{\rm i}}\,
\Psi(T_{\rm disk}-t_{\rm cool}(L,M_{\rm MS})-t_{\rm MS}(M_{\rm MS}))
\nonumber \\
&&\mbox{}\Phi(M_{\rm MS})\,\tau_{\rm cool}(L,M_{\rm MS}) \;dM_{\rm MS}
\label{wdlf}
\end{eqnarray}

\noindent  where  $\Phi(M_{\rm MS})$  is  the  initial mass  function,
$\Psi(t)$ the  star formation  rate, $t_{\rm cool}(L,M_{\rm  MS})$ the
cooling time, $t_{\rm MS}(M_{\rm  MS})$ the main sequence lifetime and
$\tau_{\rm  cool}(L,M_{\rm  MS})$  the  characteristic  cooling  time.
Moreover,  for  each  independent   realization  we  have  fitted  the
``theoretical''  white  dwarf luminosity  functions  not  only to  the
average value of the disk  white dwarf luminosity function as obtained
from our Monte  Carlo simulations but also to  the corresponding upper
and lower  values allowed by  the error bars.   In this way we  end up
with an estimate of the error in determining the disk age. The results
are shown in  table 4. As can  be seen the errors will  be small.  The
typical  error estimate  obtained  using the  actually observed  white
dwarf  luminosity function  is 1.5~Gyr,  5 times  larger.  Hence, {\sl
Gaia} will  allow a precise determination  of the age  of the Galactic
disk which  may be  compared with that  obtained using  other methods,
like turn-off stars and isochrone fitting.  In this case, moreover, it
should be taken  as well into account that {\sl  Gaia} will allow very
rigorous tests of the main sequence and red giant stellar evolutionary
models, so  additional information will be available  to constrain the
(pre-white dwarf) stellar models.

\begin{figure}
\vspace{6.5cm}
\includegraphics{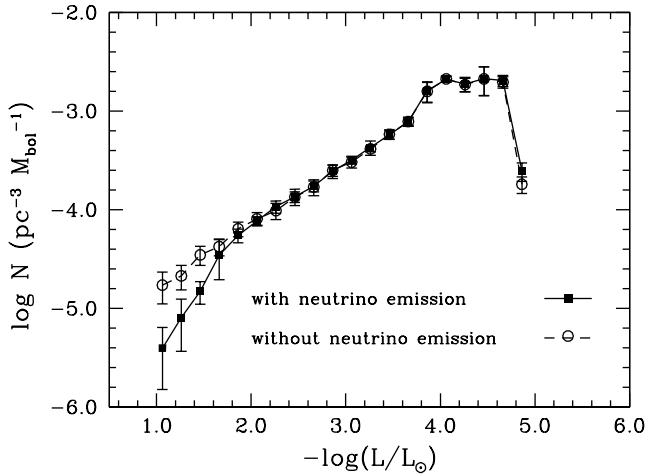}
\caption{Luminosity function  of disk white dwarfs  for two differents
         sets  of cooling  curves. The  solid squares  connected  by a
         solid line correspond to  the case in which neutrino emission
         was  properly taken  into account,  whereas the  open circles
         connected by  a dashed line  correspond to the case  in which
         neutrino emission was neglected. See text for details.}
\label{neutrinos}
\end{figure}

Next  we ask  ourselves if  {\sl Gaia}  will be  able  to discriminate
between different  cooling curves and, hence, to  place constraints on
the physical  mechanisms operating  during the cooling  process.  More
specifically, we  want to show  here that {\sl  Gaia} will be  able to
place constraints  both on the mechanisms operating  at high effective
temperatures --- basically neutrino  cooling --- and on the mechanisms
which   are   dominant    for   relatively   low   core   temperatures
(crystallization).

Neutrinos are  the dominant form of  energy loss in  model white dwarf
stars down to $\log(L/L_{\sun}) \simeq -2.0$, depending on the stellar
mass.  As  a consequence, the evolutionary timescales  of white dwarfs
at these luminosities sensitively depend  on the ratio of the neutrino
energy loss  to the photon energy  loss, and, hence, the  slope of the
white dwarf  luminosity function  directly reflects the  importance of
neutrino emission.  Although the  unified electroweak theory of lepton
interactions that is crucial for understanding neutrino production has
been  well tested  in the  high-energy regime  --- see,  for instance,
Hollik (1997) for an excellent  review --- the approach presented here
should result  in an  interesting low-energy test  of the  theory.  To
this  regard in  Fig.~\ref{neutrinos}  we show  two  disk white  dwarf
luminosity functions for which we have adopted different prescriptions
for the cooling curves. In both cases we have adopted the evolutionary
cooling sequences  of Benvenuto \&  Althaus (1999). However in  one of
the  sequences (corresponding to  open circles  connected by  a dashed
line) neutrino emission has been artificially inhibited. In passing we
also note  that, opposite to  what has been  done so far, and  for the
sake of simplicity  in this set of simulations we  have adopted a {\sl
single}  cooling sequence,  corresponding to  that of  an average-mass
$0.6\, M_{\sun}$ white dwarf made of pure oxygen. Fig.~\ref{neutrinos}
clearly shows that the drop-off in the white dwarf luminosity function
is not affected by the inclusion of the neutrino emissivity, as should
be expected  given that the  neutrino-dominated cooling phase  is very
short in  all cases.  However the  slope of the  disk white luminosity
function,  which  reflects  the  cooling  rate, is  sensitive  to  the
treatment of  neutrinos.  More interestingly, {\sl Gaia}  will be able
to measure the cooling rate and, thus, to probe the electroweak theory
at low energies.

\begin{figure}
\vspace{6.5cm}
\includegraphics{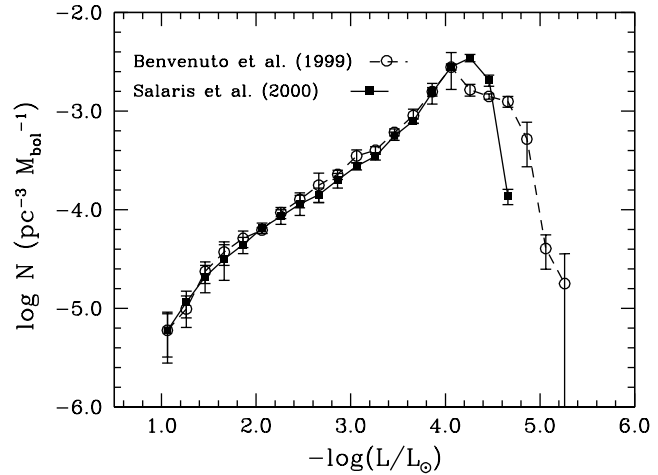}
\caption{Luminosity function  of disk  white dwarfs for  two different
         sets of  cooling sequences. The solid squares  connected by a
         solid line correspond to  the white dwarf luminosity function
         obtained when the cooling  sequences of Salaris et al. (2000)
         are used, whereas the open circles connected by a dashed line
         correspond  to the white  dwarf luminosity  function obtained
         when  the set of  cooling sequences  of Benvenuto  \& Althaus
         (1999) are used.}
\label{cooling}
\end{figure}

After  examining the  physical mechanisms  that operate  at moderately
high luminosities, say $\log(L/L_{\sun})\ga -2.0$, we focus now on one
of the crucial issues in the  theory of white dwarf cooling, namely on
crystallization and phase separation  a low core temperatures ($T_{\rm
c}\sim  10^6$~K).  In  order  to make  reliable  comparisons we  adopt
besides  our own  cooling sequences  (Salaris et  al.  2000)  those of
Benvenuto  \&  Althaus  (1999).   This  set of  cooling  sequences  is
available for  all the masses of  interest, uses a  modern equation of
state and the internal chemical profiles of Salaris et al. (1997). The
only major  difference between both  sets of cooling sequences  is the
treatment of  phase separation upon crystallization which  in the case
of the cooling sequences of  Salaris et al.  (2000) was properly taken
into account, whereas in the cooling sequences of Benvenuto \& Althaus
(1999) it was disregarded.   As discussed in Isern et  al.  (1997) the
inclusion of phase separation upon crystallization adds an extra delay
to the  cooling (and,  thus, considerably modifies  the characteristic
cooling  times at  low  luminosities), which  depends  on the  initial
chemical  profile  (the  ratio  of   carbon  to  oxygen)  and  on  the
transparency  of the  insulating  envelope. In  both  sets of  cooling
sequences the  thicknesses of the  helium buffer and of  the overlying
hydrogen envelope are the same.  Thus, the disk white dwarf luminosity
function computed  with those sets of cooling  sequences should mostly
reflect  the treatment  of  crystallization.  This  is illustrated  in
Fig.~\ref{cooling}, where  the luminosity functions  computed with the
cooling sequences of Salaris et al. (2000) --- solid squares connected
with a solid  line --- and with the cooling  sequences of Benvenuto \&
Althaus (1999) for a disk age  of 12~Gyr are displayed.  Note that for
moderately  high luminosities ---  namely $\log(L/L_{\sun})  \ga -4.0$
--- the  agreement between  both sets  of calculations  is  very good.
Obviously,  for the same  disk age,  the cut-off  of disk  white dwarf
luminosity function  computed with the cooling  sequences of Benvenuto
\& Althaus (1999) moves to lower luminosities, $\log(L/L_{\sun})\simeq
-5.0$.   Consequently,  if a  direct  measure  of  the disk  age  with
reasonable  precision is obtained  by an  independent method,  say via
turn-off  stars,  {\sl  Gaia}  will  directly  probe  the  physics  of
crystallization. It  is worth noting as  well that not  only the exact
location of the  drop-off of the disk white  dwarf luminosity function
is  affected by the  details of  the cooling  sequences but  also, the
position and  the shape of the  maximum of the  white dwarf luminosity
function, thus allowing additional tests.

\begin{figure}
\vspace{6.5cm}
\includegraphics{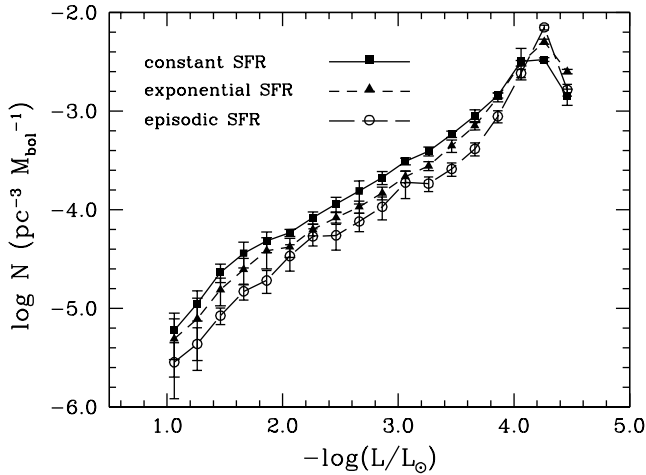}
\caption{Luminosity function of massive  disk white dwarfs for several
         star formation  histories.  The  solid line corresponds  to a
         constant volumetric star formation rate. The long dashed line
         corresponds to an episodic  star formation rate and the short
         dashed  solid line  corresponds  an exponentially  decreasing
         star formation rate.}
\label{SFR}
\end{figure}

Now  we turn  our attention  to how  the disk  white  dwarf luminosity
function may  be used  to derive the  Galactic star history.   We have
computed a  series of models in  which we have  adopted different star
formation  rates.  In  all cases  the  adopted  age  of the  disk  was
12~Gyr. For the first of our models we have adopted (as earlier in the
paper)  a constant  volumetric star  formation rate.  The  second star
formation  rate  is  on  which  is exponentially  decreasing  with  an
e-folding time $\tau=4$~Gyr.  Finally, our last adopted star formation
rate corresponds to episodic star formation, for which we have adopted
a burst of constant strength that started 1~Gyr after the formation of
the disk  and lasting for  3~Gyr.  For the  rest of the time  the star
formation activity considered  in this case was zero.  The results are
shown in  Fig.~\ref{SFR}. As  can be seen  the white  dwarf luminosity
will be  sensitive to the star formation  history. However, recovering
the exact dependence  of the star formation history  will be difficult
since the  inverse problem must be  solved (Isern et  al. 1995).  From
Eq.~(\ref{wdlf}) it  is clear  that the origin  of the problem  is the
long lifetimes  of low  mass main sequence  stars (Isern et  al. 1995;
Garc\'{\i}a--Berro  et al. 2003b).   This implies  that the  past star
formation  activity  is  still  influencing the  present  white  dwarf
birthrate. This  can be clearly  seen in Fig.~\ref{SFR}.  In  order to
solve  Eq.~(\ref{wdlf}) for the  star formation  rate there  exist two
alternatives.  The  first and most straightforward  method requires an
``a priori'' knowledge of the  shape of the star formation history and
consists  in   adopting  a   trial  function,  depending   on  several
parameters, and  search for the  values of these parameters  that best
fit  the observed  luminosity function  by minimizing  the differences
between the observational and  the computed luminosity function (Isern
et  al.  2001).  The  second  possibility  consists  in computing  the
luminosity function  of massive white  dwarfs (D\'\i az--Pinto  et al.
1994;  Isern  et  al.   1999),  which have  negligible  main  sequence
lifetimes,  thus  making  much  easier  the solution  of  the  inverse
problem.

\begin{figure}
\vspace{6.5cm}
\includegraphics{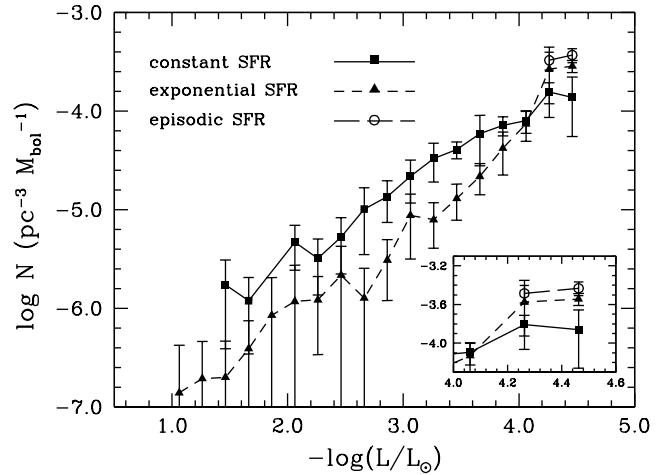}
\caption{Luminosity function of massive  disk white dwarfs for several
         star formation  histories.  The  solid line corresponds  to a
         constant volumetric star formation rate. The long dashed line
         corresponds to an episodic  star formation rate and the short
         dashed  solid line  corresponds  an exponentially  decreasing
         star formation rate. The inset shows an  expanded view of the 
         region near the maximum of the luminosity function, where the
         episodic star formation rate can be better observed.}
\label{SFRmass}
\end{figure}

In  our simulations  we obtain  a sizeable  fraction of  massive white
dwarfs,  those with  masses larger  than say  $0.8\,  M_{\sun}$, which
varies from  7\% for the constant  star formation history,  to 4\% for
the  exponential  one and  to  3\%  for  the episodic  star  formation
history. These fractions are enough  to obtain the history of the star
formation  activity  in the  solar  neighborhood  (D\'\i az--Pinto  et
al. 1994). Although these fractions  may seem small when taken at face
value, the  absolute numbers of  massive white dwarfs  are impressive,
since  for the case  of a  constant star  formation rate,  700 massive
white  are  expected to  be  found, whereas  for  the  other two  star
formation histories  500 and  300 massive white  dwarfs will  be found
respectively, thus allowing a determination of the luminosity function
of {\sl massive} white dwarfs.  Such luminosity functions are shown in
Fig.~\ref{SFRmass} for  the three  star formation rates  studied here.
As this  figure clearly  shows we  will be able  to obtain  a reliable
determination of  the star formation  rate. It is interesting  to note
that the slopes  of the luminosity functions computed  with a constant
and  an  exponentially  decreasing  star  formation  rates  are  quite
different now, in contrast  with the behavior shown in Fig.~\ref{SFR}.
Moreover  the contribution  of  the episodic  star  formation rate  is
concentrated in the very last luminosity bins.

\subsection{The halo white dwarf luminosity function}

\begin{figure}
\vspace{6.5cm}
\includegraphics{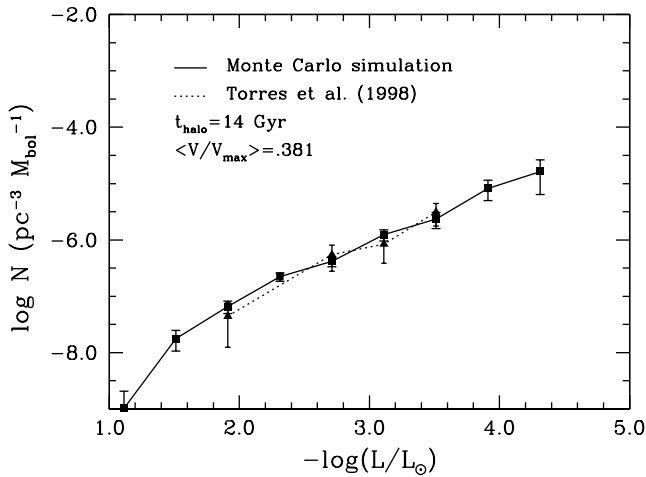}
\caption{Luminosity  function of  halo  white dwarfs.  The solid  line
         corresponds  to the  simulations presented  here,  assuming a
         recovery franction of 50\%. The dashed line is the luminosity
         function  of  Torres et  al.  (1998).  See  text for  further
         details.}
\label{flhalo}
\end{figure}

\begin{figure}
\vspace{6.5cm}
\includegraphics{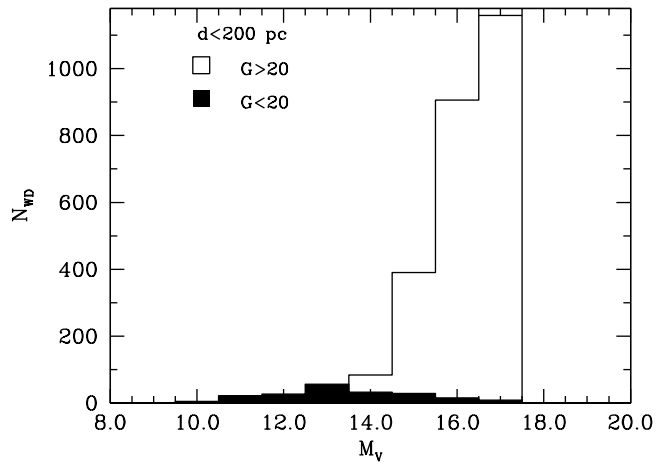}
\caption{Distribution  of halo  white  dwarfs detected  by {\sl  Gaia}
         (solid  histogram)  compared   to  the  distribution  of  the
         simulated sample of halo  white dwarfs (empty histogram), for
         an halo age of 14~Gyr, and a recovery fraction of 50\%.}
\label{hmvhalo}
\end{figure}

In figure~\ref{flhalo}  we show the luminosity function  of halo white
dwarfs that  {\sl Gaia}  will observe  for a halo  age of  14~Gyr, and
assuming that only 50\% of the halo white dwarfs with $H<18$ and $V-I<
0.3$ are correctly classified as halo white dwarfs.  As borne out from
Fig.~\ref{flhalo}, {\sl Gaia} will be  able to measure only the bright
portion of the  white dwarf luminosity function of  halo white dwarfs,
which carries  valuable information  about the initial  mass function.
However, the cut-off of the  luminosity function --- which provides an
independent estimate  of the age of  the stellar halo ---  will not be
detected.   This is obviously  due to  the cut  in magnitudes  of {\sl
Gaia},  which  will  be  $G\sim  20^{\rm  mag}$.   According  to  this
magnitude cut  and given that the  age of the stellar  halo is $t_{\rm
halo}\ga 13$~Gyr,  a considerable fraction  of halo white  dwarfs will
not be detected by {\sl Gaia}. This is assessed in Fig.~\ref{hmvhalo},
where  the  distribution  of   white  dwarfs  with  $G<20$  ---  those
detectable by  {\sl Gaia} --- is  compared to the  total population of
halo  white  dwarfs.   Clearly,   very  few  halo  white  dwarfs  with
magnitudes  close to  that of  the cut-off  will be  observed  by {\sl
Gaia}, thus preventing us to  directly measure the age of the Galactic
halo.

One important  concern is whether  or not the  advanced classification
techniques mentioned above are mandatory in order to obtain a reliable
luminosity function.  In  order to check this issue  we have proceeded
as  follows. First we  have assumed  that we  are able  to distinguish
between halo and  disk candidates with a success  rate of 100\%. After
this, in a  second set of calculations we have  assumed that only 50\%
of halo white  dwarfs are correctly recovered in  the region delimited
by $H<18$ and  $V-I< 0.3$. Finally, in a third  set of calculations we
have adopted a recovery fraction  of 25\% within this region. The halo
white  dwarf  luminosity  function  presented  in  figure~\ref{flhalo}
corresponds  to the  second case.   However,  we have  found that  the
results  are relatively  insensitive  to the  recovery fraction.   The
reason for this  behaviour is easy to understand:  the region in which
disk  and  halo  white  dwarfs  are not  well  identified  corresponds
precisely to  the brightest white  dwarfs and, hence,  to luminosities
for which the density of white dwarfs per unit bolometric magnitude is
very small.  Consequently, although {\sl Gaia} will only determine the
bright portion of the halo white dwarf luminosity function, there will
not  be serious systematic  errors for  luminosities $\log(L/L_{\sun})
\ga -2.0$.

\begin{figure}
\vspace{6.5cm}
\includegraphics{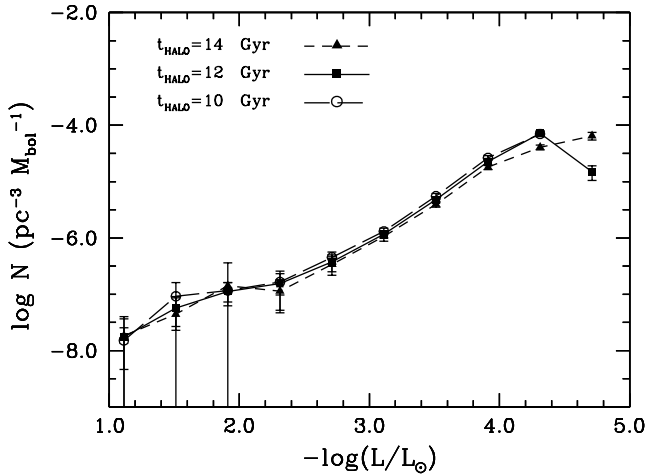}
\caption{White  dwarf  luminosity  functions  for several  halo  ages,
         ranging from  10 to 14~Gyr,  assuming a recovery  fraction of
         50\%.}
\label{thalo}
\end{figure}

As we  have shown, it  is not expected  that {\sl Gaia}  will directly
measure the  age of the Galactic  halo by finding white  dwarfs at the
end  of the  halo white  dwarf cooling  sequence.  However,  one could
imagine that the halo age  could be still be somehow constrained since
a younger  halo could  generate stars above  the {\sl  Gaia} magnitude
limit.  In  Fig.~\ref{thalo} we  explore such possibility  by adopting
several halo  ages and, again,  assuming a recovery fraction  of 50\%.
This figure  clearly shows that for  reasonable halo ages  this is not
the case, since the drop-off of the white dwarf luminosity function is
located at luminosities well beyond the capabilities of {\sl Gaia}, as
anticipated in Isern et al. (1998a).

\begin{figure}
\vspace{6.5cm}
\includegraphics{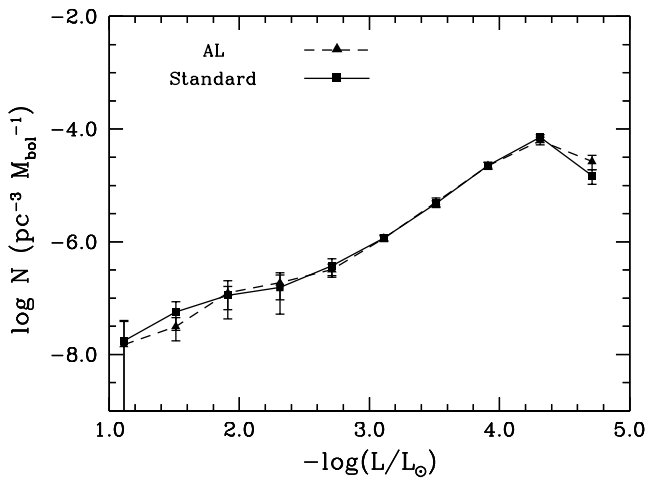}
\caption{Halo white  dwarf luminosity  functions for two  initial mass
         functions, adopting a halo age of 12~Gyr and an age spread of
         1~Gyr, again assuming a recovery fraction of 50\%.}
\label{IMFhalo}
\end{figure}

In  Fig.~\ref{IMFhalo} we explore  the sensitivity  of the  halo white
dwarf  luminosity   function  to  the  choice  of   the  initial  mass
function. As we have done so far  we have adopted a halo age of 12~Gyr
and an  age spread of 1~Gyr.   We have simulated two  halo white dwarf
populations, the first  one according to the initial  mass function of
Adams \&  Laughlin (1996) and the  second one to  our standard initial
mass function (Scalo 1998).  Unfortunately {\sl Gaia} will not be able
to  distinguish  between  these  initial  mass  functions,  as  figure
\ref{IMFhalo} clearly shows.

This is not surprising at all.  Assuming that the halo was formed in a
burst of  star formation  of negligible duration,  then for  all white
dwarfs we have:

\begin{equation}
t_{\rm HALO}\simeq t_{\rm MS}(M_{\rm MS})+t_{\rm cool}(L,M_{\rm MS})
\end{equation}

This  means that,  given an  age of  the halo,  $t_{\rm  HALO}$, there
exists a function $M_{\rm MS}=M_{\rm  MS}(L)$ or, in other words, that
the  white dwarfs  contributing to  each luminosity  bin of  the white
dwarf luminosity function have the same mass. Taking this into account
we have:

\begin{equation}
n(L)\simeq\frac{dn}{dM_{\rm MS}}\frac{dM_{\rm MS}}{dL}=
\Phi(M_{\rm MS})\frac{dM_{\rm MS}}{dL}
\end{equation}

\begin{figure}
\vspace{6.5cm}
\includegraphics{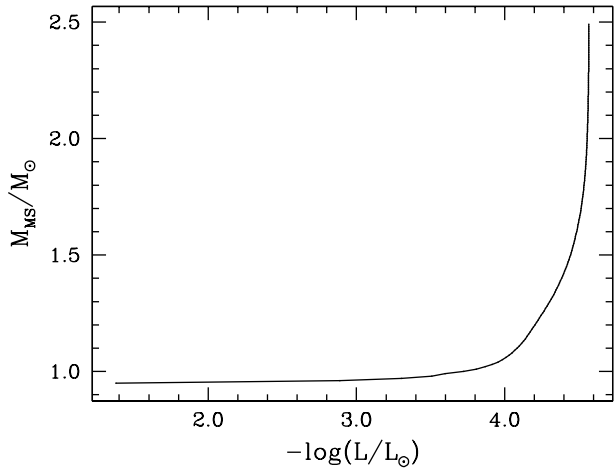}
\caption{Mass in the main sequence contributing to each luminosity bin
         for a burst of  star formation  of negligible  duration which
         happened 12~Gyr ago.}
\label{lms}
\end{figure}

The first term in this  equation is the initial mass function, whereas
the second term is related  to the cooling times. Hence, provided that
we have  reliable characteristic cooling  times, the halo  white dwarf
luminosity  function  could  be  eventually used  to  retrieve  useful
information about the initial mass  function of the halo, if different
from that  of the disk  populations. However, most of  the information
regarding the initial mass function concentrates in the low luminosity
bins. In  particular, in  Fig.~\ref{lms} we show  the the mass  of the
main sequence  which contributes  to each luminosity  bin of  the halo
white dwarf luminosity function. As  can be seen there $M_{\rm MS}(L)$
remains   almost   flat  up   to   luminosities   of   the  order   of
$\log(L/L_{\sun})\simeq  -3.5$, and  then  a very  steep  rise can  be
observed.   As we  have already  discussed, the  number of  halo white
dwarfs of  these very low  luminosities that {\sl Gaia}  will probably
observe is small and, consequently, will not allow us to draw definite
conclusions about the shape of the initial mass function.

\begin{figure}
\vspace{6.5cm}
\includegraphics{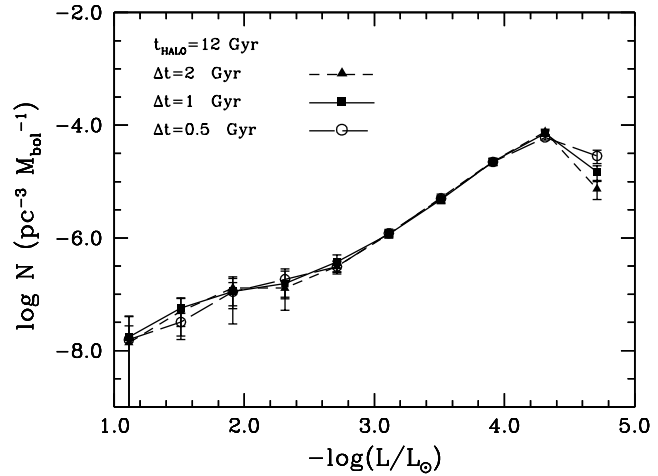}
\caption{White  dwarf  luminosity  functions  for several age spreads
         of  the Galactic halo  ranging  from 0.5  to 2.0~Gyr,  again
         assuming a recovery  fraction of 50\%.}
\label{Dthalo}
\end{figure}

Finally, in Fig.~\ref{Dthalo} we explore the effects in the luminosity
function of halo white dwarfs  of different age spreads of the adopted
burst of star formation. We have adopted a halo age of 12~Gyr, whereas
the durations  of the star  formation bursts were 2~Gyr  (short dashed
line and solid  triangles), 1~Gyr (solid line and  filled squares) and
0.5~Gyr   (long  dashed  line   and  open   circles).   As   shown  in
Fig.~\ref{Dthalo} the three  curves are almost indistinguishable, thus
preventing  us from obtaining  a better  understanding the  process of
formation of  the Galactic  halo by using  the luminosity  function of
halo  white dwarfs.  The  reason is  quite simple  and related  to the
behavior of $M_{\rm MS}(L)$ as shown in Fig.~\ref{lms}. In particular,
we have that for the duration  of the bursts of star formation adopted
here (2, 1  and 0.5~Gyr) --- which we believe  cover a realistic range
of age spreads  --- the corresponding masses of  the white dwarfs just
entering  into the  cooling phase  are, respectively,  0.57,  0.59 and
$0.61\,  M_{\sun}$, and thus  their respective  characteristic cooling
times, $\tau_{\rm cool}$, are  very similar.  Since at luminosities of
$\log(L/L_{\sun})\approx  -2$ the function  $M_{\rm MS}(L)$  is almost
flat, the bright  branch of the luminosity function  only reflects the
speed  of  cooling,  washing  out  any  other  information  (Isern  et
al. 1998a).

\section{Conclusions and discussion}

In this paper we have explored  the impact {\sl Gaia} will have on our
understanding of  the Galactic white  dwarf population. We  have shown
that the superb astrometric capabilities of {\sl Gaia} will provide us
with  an   unprecedented  number   of  white  dwarfs   with  excellent
astrometric measurements.  In particular we  have shown that  the disk
white dwarf population will be  probed up to distances of 400~pc, with
typical errors smaller  than 10\%, both in proper  motion and parallax
and with a  completeness ranging from nearly 100\%  for objects within
100~pc  to 30\% for  objects within  400~pc, when  a magnitude  cut of
$G_{\rm cut}=21$  is adopted.  The performances, of  course, are worse
for a magnitude cut of 21.   Thus, {\sl Gaia} will determine with high
accuracy the  disk white dwarf  luminosity function and  its drop-off.
We have also shown that  this excellent situation will not pertain for
the  halo  white  dwarf   population.   In  particular,  although  the
astrometric  measurements  will  be   highly  accurate  as  well,  the
completeness of the survey will be much smaller, typically 50\% within
100~pc.  We  have also  analyzed how to  distinguish between  halo and
disk white dwarfs  and we have demonstrated that  although the reduced
proper motion  diagram will be of  some help in  this regard, advanced
classification  techniques will  be  required to  extract the  maximum
amount of information from  the halo white dwarf population.  Finally,
we have  also studied what  would be the  typical disk and  halo white
dwarf luminosity functions that {\sl Gaia} will eventually obtain, and
we have  analyzed what  could be the  attainable scientific  goals. We
have  found  that  the  disk  white  dwarf  luminosity  function  will
constrain the age  of the Galactic disk with  good accuracy, using the
observed drop-off  in the disk  white dwarf luminosity  function, very
much  improving the present  day constrains.  {\sl Gaia}  will provide
very precise information  on the physical mechanisms (crystallization,
phase  separation,  etc)  operating  during  the  cooling  process  by
comparing the  theoretical luminosity  functions of disk  white dwarfs
with the  observations. The luminosity function of  massive disk white
dwarfs will constrain the star formation history of the Galactic disk,
whereas the lower  mass white dwarfs will offer  few constraints.  For
the halo we have found that only the bright portion will be accessible
to {\sl  Gaia}, thus preventing  us from getting  valuable information
about the initial mass function of  the Galactic halo, or even its age
or the duration  of hypothetical burst of star  formation which led to
its formation (at least from its white dwarfs).

White dwarfs are well studied  objects and the physical processes that
control their  evolution are reasonably  well understood ---  see, for
instance, the reviews of Isern  et al. (1998b), Koester (2002), Hansen
\& Liebert (2003)  and Isern \& Garc\'\i a--Berro  (2004) --- at least
up   to   moderately   low   luminosities   ---  of   the   order   of
$\log(L/L_{\sun})=  -3.5$.   In  fact,  most  phases  of  white  dwarf
evolution can be succesfully  characterized as a cooling process. That
is,  white  dwarfs slowly  radiate  at  the  expense of  the  residual
gravothermal energy.  The  release of this energy lasts  for long time
scales (of  the order of the  age of the Galactic  disk $\sim 10^{10}$
yr).  The  mechanical structure  of white dwarfs  is supported  by the
pressure  of the gas  of degenerate  electrons, whereas  the partially
degenerate  outer   layers  control  the  flow   of  energy.   Precise
spectrophotometric data  --- like those  that {\sl Gaia}  will provide
--- would certainly  introduce very  tight constraints on  the models.
Specifically,  {\sl  Gaia} will  allow  us  to  test the  mass--radius
relationship, which  is still today  not particularly well  tested, by
analyzing  the  spectrophotometric data  of  white  dwarfs with  known
parallaxes ---  which will  be of the  order of several  hundreds (see
table  3).  By  comparing  the theoretical  models  with the  observed
properties  of white dwarfs  belonging to  binary systems,  {\sl Gaia}
will also  be able to constrain  the relation between the  mass in the
main sequence and the mass of the resulting white dwarf.

Given their long cooling timescales,  white dwarfs have been used as a
tool  to extract  useful information  about  the past  history of  our
Galaxy.  The large number of white dwarfs that {\sl Gaia} will observe
will allow us  to probe the structure and dynamics of  the Galaxy as a
whole, tracing back a hypothetical merger episode in the Galactic disk
(Torres et al.   2001). Moreover, it will provide  new clues about the
halo white dwarf population and its contribution to the mass budget of
our Galaxy --- see, for instance,  Isern et al. (1998a), Torres et al.
(2002) and Garc\'\i  a--Berro et al.  (2004),  and references therein.
Additionally, the  disk white dwarf luminosity function  has been used
to derive  constraints on the  rate of variation of  the gravitational
constant (Garc\'\i a--Berro et al.   1995). The accuracy of this bound
is mainly  limited by  the statistical significance  of the  very last
bins  of the  white dwarf  luminosity  function. Given  the huge  step
forward that  {\sl Gaia} will introduce  in the number  counts of disk
white dwarfs it is foreseeable that  a very tight upper limit on $\dot
G/G$ will become eventually available.  This, in turn, will pose tight
constraints    in    current    theories   with    extra    dimensions
(Lor\'en--Aguilar et al. 2003).

In summary, in this work we have shown how an astrometric mission like
{\sl Gaia} could dramatically increase the number of white dwarfs accessible
to  good  quality observations.   The  increase  of the  observational
database  will   undoubtely  have  a  large  impact   in  our  current
understanding of the history and structure of the Galaxy as well as on
the theoretical  models of white  dwarf cooling, which, in  turn, will
certainly  influence our knowledge  of the  physics of  dense plasmas.
Nevertheless,   follow-up   ground-based   observations,   theoretical
improvements and advanced classification  methods (Torres et al. 1998;
Garc\'\i a--Berro et al. 2003a) will be needed in order to analyze the
disk and halo populations.

\vspace{0.5 cm}

\noindent {\sl Acknowledgements.}  Part  of this work was supported by
the MCYT grants  AYA04094--C03-01 and 02, by the  European Union FEDER
funds, and by  the CIRIT. L.G. Althaus is  gratefully acknowledged for
stimutaling  discussions  and  for  providing  the  cooling  sequences
necessary for  computing the disk white dwarf  luminosity functions of
Fig.~\ref{neutrinos}.  We  also acknowledge our  referee, Chris Flynn,
for  a very careful  reading of  the original  manuscript and  for his
valuable and constructive criticisms and comments.


\begin{thebibliography}{}

\bibitem{AL96}    Adams, F.C., \& Laughlin, G., 1996, ApJ, 468, 586
\bibitem{AEA96}   Alcock, C.,  Allsman, R.A.,  Axelrod, T.S., Bennett,
                  D.P., Cook,  K.H., Freeman, K.C., Griest, K., Guern,
                  J.A.,  Lehner,  M.J., Marshall, S.L.,  Park,  H.-S.,
                  Perlmutter, S., Peterson, B.A., Pratt,  M.R., Quinn,
                  P.J.,  Rodgers, A.W.,  Stubbs, C.W., Sutherland, W.,
                  1996, \apj, 461, 84
\bibitem{AEA00}   Alcock,  C., Allsman,  R.A.,  Alves, D.R.,  Axelrod, 
                  T.S.,  Becker,  A.C.,  Bennett,  D.P., Cook,   K.H., 
                  Dalal,  N., Drake,  A.J., Freeman,  K.C., Geha,  M.,  
                  Griest, K., Lehner,  M.J., Marshall,  S.L., Minniti,  
                  D.,  Nelson,  C.A.,  Peterson,  B.A.,  Popowski, P.,  
                  Pratt,  M.R., Quinn, P.J., Stubbs, C.W., Sutherland,  
                  W., Tomaney, A.B., Vandehei, T., \& Welch, D., 2000,
                  \apj, 542, 281
\bibitem{BA99}    Benvenuto, O.G., \& Althaus, L.G., 1999, MNRAS, 303,
                  30
\bibitem{B87}     Binney,  J., \&  Tremaine, H., 1987, {\sl  ``Galactic
		  Dynamics''} (Princeton Univ.  Press: Princeton)
\bibitem{D97}     Dehnen, W., \& Binney, J.J., 1997, MNRAS, 287, L5
\bibitem{DEA94}   D\'\i az--Pinto, A., Garc\'\i a--Berro,  E., Hernanz, 
                  M.,  Isern, J., \&  Mochkovitch, R., 1994, A\&A, 282, 
                  86
\bibitem{GB88b}   Garc\'\i a--Berro, E., Hernanz, M., Mochkovitch, R.,
		  \& Isern, J., 1988, \aap, 193, 141
\bibitem{GBEA95}  Garc\'\i a--Berro, E., Hernanz, M.,  Isern, J.,   \& 
                  Mochkovitch, R., 1995, MNRAS, 277, 801
\bibitem{GBEA99}  Garc\'\i a--Berro,  E.,  Torres, S.,  Isern, J.,  \&
                  Burkert, A., 1999, MNRAS, 302, 173
\bibitem{GBEA03}  Garc\'\i a--Berro,  E.,  Torres, S.,  \&  Isern, J.,
                  2003a, Neural Networks, 16, 405
\bibitem{GB3b}    Garc\'\i a--Berro, E.,   Torres, S.,  \&  Isern, J., 
                  2003b,  in  {\sl  ``White Dwarfs''},  Eds.:   D.  de 
                  Martino,  R. Silvotti,  J.E. Solheim \&  R. Kalytis, 
                  NATO  Science  Series (Kluwer  Academic  Publishers: 
                  Dordrecht)  vol. 105, 23 
\bibitem{GBEA04}  Garc\'\i a--Berro,  E.,  Torres, S.,  Isern, J.,  \&
                  Burkert, A., 2004, A\&A, 418, 53
\bibitem{HL97}    Hakkila,  J., Myers,  J.M., Stidham, B.J., Hartmann,
                  D.H., 1997, AJ, 114, 2043
\bibitem{HL03}    Hansen,  B.M.S., \&  Liebert, J.W.,  2003, Ann. Rev. 
                  A\&A, 41, 465
\bibitem{HA94}    Hernanz,  M.,  Garc\'\i  a--Berro,  E.,  Isern,  J.,
		  Mochkovitch,  R.,  Segretain,  L., \&  Chabrier, G.,
		  1994, \apj, 434, 652
\bibitem{H97}     Hollik, W., 1997, J. Phys. G: Nucl. Part. Phys., 23, 
                  1503
\bibitem{IL89}    Iben, I. Jr., Laughlin, G., 1989, ApJ, 341, 312
\bibitem{I95}     Isern,  J.,  Garc\'\i  a--Berro,  E.,  Hernanz,  M., 
                  Mochkovitch,  R.,  \&  Burkert,  A.,  1995,  in {\sl 
                  ``White Dwarfs''},  Eds.:  D. Koester  \&  K. Werner
                  (Berlin: Springer Verlag), 19
\bibitem{IEA97}   Isern, J.,  Mochkovitch, R.,  Garc\'\i a--Berro, E., 
                  \& Hernanz, M., 1997, \apj, 485, 308
\bibitem{I98a}    Isern,  J.,  Garc\'\i  a--Berro,  E.,  Hernanz,  M.,
                  Mochkovitch, R., \& Torres, S., 1998a, \apj, 503,
\bibitem{I98b}    Isern,  J., Garc\'\i  a--Berro, E.,  Hernanz, M., \&
                  Mochkovitch,  R., 1998b,  Jour.  of Phys.: Condensed 
                  Matter, 10, 11263
\bibitem{I99}     Isern,  J.,  Hernanz,  M.,  Garc\'\i  a--Berro,  E., 
                  \& Mochkovitch, R., 1999, in  {\sl  ``Proc.  of  the 
                  11$^{\rm th}$ Workshop on  White Dwarfs},  ASP Conf. 
                  Ser., vol.  169,  Eds.: J.E. Solheim \& E.G. Meistas 
                  (ASP: San Francisco), 408
\bibitem{I01}     Isern, J.,  Garc\'\i a--Berro,  E.,  \& Salaris, M., 
                  2001, in {\sl ``Astrophysical Ages and  Timescales},
                  ASP Conf. Ser.,  vol. 245,  Eds.: T. von Hippel,  C.
                  Simpson \& N. Manset (ASP: San Francisco), 328
\bibitem{IGB04}   Isern, J., \& Garc\'\i a--Berro,  E., 2004,  in {\sl
                  Lecture Notes  and Essays in  Astrophysics I}, Eds.:
                  A.  Ulla  \& M. Manteiga (Publications  of the RSEF:
                  Vigo),                    23,                   {\tt
                  http://www.slac.stanford.edu/econf/C0307073}
\bibitem{J90}     James, F., 1990, Comput.  Phys.  Commun., 60, 329
\bibitem{K02}     Koester, D., 2002, A\&A Rev., 11, 33
\bibitem{LEA01}   Lasserre,  T.,  Afonso,  C., Albert, J.N., Andersen, 
                  J., Ansari, R., Aubourg, {\'E}., Bareyre, P., Bauer, 
                  F., Beaulieu,  J.P., Blanc,  G., Bouquet,  A., Char, 
                  S., Charlot,  X., Couchot, F.,  Coutures, C., Derue, 
                  F.,  Ferlet,   R., Glicenstein,  J.F., Goldman,  B., 
                  Gould,  A.,  Graff,  D., Gros,  M., Haissinski,  J.,
                  Hamilton,  J.C.,  Hardin,  D., de Kat, J., Kim,  A., 
                  Lesquoy, {\'E}., Loup, C., Magneville, C.,  Mansoux,   
                  B., Marquette, J.B., Maurice, {\'E}., Milsztajn, A., 
                  Moniez, M., Palanque-Delabrouille, N., Perdereau, O.,
 		  Pr{\' e}vot, L., Regnault, N., Rich,  J., Spiro, M.,
		  Vidal-Madjar,  A., Vigroux,  L., \& Zylberajch,  S.,
		  2001, \aap, 355, L39
\bibitem{LEA03}   Lor\'en--Aguilar, P., Garc\'\i a--Berro,  E., Isern,
                  J., \& Kubyshin, Y.A.,  Class. \& Quantum Grav., 20,
                  3885
\bibitem{MSL97}   Markovi\'c, D., \&  Sommer-Larsen, J., 1997,  MNRAS,
		  288, 733
\bibitem{M81}     Mihalas, D., \&  Binney, J.,  1981, {\sl  ``Galactic
                  Astronomy''} (W.H. Freeman \& Co.: New York)
\bibitem{NS90}    Noh, H.-R., \& Scalo, J., 1990, \apj, 352, 605
\bibitem{O65}     Ogorodnikov, K.F., 1965, {\sl ``Dynamics of  Stellar
                  Systems''} (Pergamon Press: New York)
\bibitem{OS96}    Oswalt,  T.D.,  Smith, J.A., Wood, M.A., \& Hintzen, 
                  P., 1996, Nature, 382, 692
\bibitem{P02}     Perryman, M.A.C.,  2002, in  EAS  Pub. Ser., Vol. 2,
                  Proc.  of  {\sl ``GAIA: A European Space  Project''}
                  (EDP Sciences: Les Ulis, France)
\bibitem{PEA01}   Perryman, M.A.C., de Boer, K.S., Gilmore,  G., Hoeg,
                  E.,   Lattanzi,  M.G.,  Lindegren,   L.,  Luri,  X., 
                  Mignard,  F., Pace, O.,  de Zeeuw, P.T., 2001, A\&A, 
                  369, 339
\bibitem{P86}     Press,  W.H.,  Flannery,  B.P.,   Teukolsky,   S.A.,
                  Vetterling, W.T., 1986, {\sl ``Numerical  Recipes''}
                  (Cambridge Univ. Press; Cambridge, UK)
\bibitem{R00}     Richer, H.B.,  Hansen, B., Limongi, M,, Chieffi, A.,
                  Straniero,  O., \&  Fahlman, G.G., 2000,  \apj, 529, 
                  318
\bibitem{SEA97}   Salaris,  M., Dom\'\i nguez,  I., Garc\'\i a--Berro,
                  E.,  Hernanz,  M.,  Isern,  J., \&  Mochkovitch, R., 
                  1997, \apj, 486, 413
\bibitem{SEA00}   Salaris,  M., Garc\'\i  a--Berro, E.,  Hernanz,  M.,
                  Isern, J.  \& Saumon, D., 2000, ApJ, 544, 1036
\bibitem{SG03}    Salim, S., \& Gould, A., 2003,  ApJ, 582, 1011
\bibitem{S98}     Scalo, J., 1998, in {\sl ``The Stellar  Initial Mass
		  Function''}, Eds.:  G.  Gilmore \&  D.  Howell (PASP
		  Conference Series: San Francisco), Vol.  142, 201
\bibitem{S68}     Schmidt, M., 1968, \apj, 151, 393
\bibitem{TEA98}   Torres,  S., Garc\'\i  a--Berro,  E., \& Isern,  J.,
                  1998, ApJ, 508, L71
\bibitem{TEA01}   Torres, S.,  Garc\'\i a--Berro, E., Burkert,  A., \& 
                  Isern, J., 2001, MNRAS, 328, 492
\bibitem{TEA02}   Torres, S.,  Garc\'\i a--Berro, E., Burkert,  A., \& 
                  Isern, J., 2002, MNRAS, 336, 971
\bibitem{WA87}    Winget, D.E.,  Hansen, C.J., Liebert, J., van  Horn,
		  H.M., Fontaine, G., Nather,  R.E., Kepler, S.O., \&
		  Lamb, D.Q., 1987, \apj, 315, L77

\end{thebibliography}
\end{document}